\documentclass[11pt]{article}
\usepackage{amsmath}
\usepackage{amsfonts}
\usepackage{amssymb}
\usepackage{graphicx}

\def\bea{\begin{eqnarray}}
\def\eea{\end{eqnarray}}

\begin{document}
\begin{center}
\LARGE {  Correspondence between the contracted BTZ solution of
cosmological topological massive gravity and two-dimensional
Galilean conformal algebra }
\end{center}
\begin{center}
{\bf M. R Setare\footnote{rezakord@ipm.ir} \\  V. Kamali\footnote{vkamali1362@gmail.com}}\\
 {\ Department of Science, Payame Noor University, Bijar, Iran}
 \\
 \end{center}
\vskip 3cm
\begin{abstract}
We show that BTZ black hole solution of Cosmological Topological
Massive Gravity (CTMG) have a hidden conformal symmetry. In this
regard, we consider the wave equation of a massless scalar field
propagating in BTZ spacetime and find the wave equation
 could be written in terms of the $SL(2,R)$
quadratic Casimir. From the conformal coordinates, the
temperatures of the dual CFTs could be read directly. Moreover,
we compute the microscopic entropy of the dual CFT by Cardy
formula and find a perfect match to Bekenstein-Hawking entropy of
BTZ black hole. Then we consider Glilean conformal algebras (GCA),
which arises as a contraction of relativistic conformal
algebras~~($x\rightarrow \epsilon x$, ~~ $t\rightarrow
t$,~~$\epsilon \rightarrow 0$). We show that there is a
correspondence between $GCA_2$ on the boundary and contracted BTZ
in the bulk. For this purpose we obtain the central charges and
temperatures of $GCA_2$. Then we compute the microscopic entropy
of the $GCA_2$ by Cardy formula and find a perfect match to
Bekenstein-Hawking entropy of BTZ black hole in non-relativistic
limit. The absorption cross section of a near region scalar field
also matches to microscopic absorption cross section of the dual
$GCA_2$. So we find further evidence that show correspondence
between contracted BTZ black hole and 2-dimensional Galilean
conformal algebra.

\end{abstract}
\newpage
\section{Introduction}
Recently, there has been some interest in extending the AdS/CFT
correspondence to non-relativistic field theories \cite{a},
\cite{b}. The Kaluza-Klein type framework for non-relativistic
symmetries, used in Refs. \cite{a}, \cite{ b}, is basically
identical to the one introduced in \cite{d} (see also \cite{e}).
The study of a different non-relativistic limit was initiated in
\cite{f}, where the non-relativistic conformal symmetry obtained
by a parametric contraction of the relativistic conformal group.
Galilean conformal algebra (GCA) arises as a contraction
relativistic conformal algebras \cite{f}, \cite{c}, where in
$3+1$ space-time dimensions the Galilean conformal group is a
fifteen parameter group which contains the ten parameter Galilean
subgroup. Infinite dimensional  Galilean conformal group has been
reported  in \cite{c}, the generators of this group are :
~$L^n=-(n+1)t^n
x_i\partial_i-t^{n+1}\partial_t$,~$M^{n}_i=t^{n+1}\partial_i$
and~$J_{ij}^n=-t^n(x_i\partial_j-x_j\partial_i)$~ for an arbitrary
integer $n$, where  i and  j are specified by the spatial
directions. There is a finite dimensional subgroup of the
infinite dimensional Galilean conformal group which generated by
~($J_{ij}^0, L^{\pm 1}, L^0, M_i^{\pm 1}, M_i^0$). These
generators are obtained by contraction ($~t\rightarrow t, ~x_i
\rightarrow \epsilon  x_i, ~\epsilon \rightarrow 0,~ v_i \sim
\epsilon$ ) of the relativistic conformal generators. Recently
the authors of \cite{1} (see also \cite{bag1}) have shown that the
$GCA_2$ is the asymptotic symmetry of CTMG in the
non-relativistic limit. They have obtained the central charges of
$GCA_2,$ and also a non-relativistic generalization of Cardy
formula. In the present paper we obtain similar result by another
method. But our aim in this paper is more than this. We show that
the BTZ black hole solution of CTMG have a hidden conformal
symmetry, not only in the relativistic case, but also in the
non-relativistic case. We show that for massless scalar field, in
non-relativistic CTMG there exist a finite Galilean conformal
symmetry acting on the solution space. According to our knowledge
this is the first paper that study the hidden conformal symmetry
for a black hole
in the non-relativistic case.\\

Recent investigation on the holographic dual descriptions for the
black holes have achieved substantial success. According to the
Kerr/CFT correspondence \cite{11}, the microscopic entropy of
four-dimensional extremal Kerr black hole was calculated by
studying the dual chiral conformal field theory associated with
the diffeomorphisms of near horizon geometry of Kerr black hole.
Subsequently, this work was extended to the case of near-extreme
black holes \cite{21}. The main progress are made essentially on
the extremal and near extremal limits in which the black hole
near horizon geometries consist a certain AdS structure and the
central charges of dual CFT can be obtained by analyzing the
asymptotic symmetry following the method in \cite{4} or by
calculating the boundary stress tensor of the $2D$ effective
action \cite{5}. Recently, Castro, Maloney and Strominger
\cite{61} have given evidence that the physics of Kerr black
holes might be captured by a conformal field theory. The authors
have discussed that the existence of conformal invariance in a
near horizon geometry is not necessary  condition, instead the
existence of a local conformal invariance in the solution space
of the wave equation for the propagating field is sufficient to
ensure a dual CFT description(see also \cite{71}). The scalar
Laplacian in the low frequency limit could be written as the
$SL(2,R)$ quadratic Casimir, showing a hidden $SL(2,R) \times
SL(2,R)$ symmetries. In the microscopic description, using the
Cardy formula for the microscopic degeneracy,
they reobtain the Bekenestein-Hawking entropy of the black hole.\\
In the present paper we investigate the massless scalar wave
equation in the background of BTZ black hole solution of CTMG and
show the wave equation can be written in terms of $SL(2,R)$
Casimir invariants. From the conformal coordinates introduced in
\cite{11}, we read the temperature of the dual CFT. The
microscopic counting  support this holographic picture. Then we
consider the non-relativistic limit of both side of this
correspondence, i.e, the non-relativistic limit of $2D$ CFT which
give us $GCA_2$ from one side, and non-relativistic limit of CTMG
which give us contracted BTZ from another side. For this purpose
we obtain the central charges and temperatures of $GCA_2$. Then we
showe that the radial part of Klein-Gordon (KG) equation in the
background of contracted BTZ black hole, where ($j\rightarrow
\epsilon j$, $\varphi\rightarrow \epsilon \varphi$), can be given
by the non-realtivistic limit of quadratic Casimir of $SL(2,R)$.
We could read the $GCA_2$ temperatures from the correspondence of
radial part of non-relativistic KG equation and Casimir of
$GCA_2$. Then we compute the microscopic entropy of the $GCA_2$ by
Cardy formula and find a perfect match to Bekenstein-Hawking
entropy of contracted BTZ black hole. After that in section 4, we
compute the absorption cross section of a near region scalar
field and find perfect match to the microscopic cross section in
dual $GCA_2$. These results supports the idea of a correspondence
between contracted BTZ black hole and dual Galilean conformal
algebra in $2-$dimension.

\section{Massless Scalar field in the background of CTMG }
In this section we introduced the idea of the hidden conformal
symmetry into the CTMG, and obtain the Virasoro algebras as a
local symmetry of massless scalar fields propagating in the BTZ
black hole background. The existence of the 2-dim CFT behind the
asymptotically $AdS_3$ spacetime, including the BTZ black hole
solutions, was already pointed out by using the Brown-Henneaux's
method, see \cite{4} for the case without the gravitational
Chern-Simons term, and \cite{n1} for the case with the Chern-
Simons term. \\
We show that for massless scalar field $\Phi$ propagating in the
background of cosmological topological massive gravity (CTMG),
there exist a $SL(2,R)_{L}\times SL(2,R)_{R}$ conformal symmetry
acting on the solution space. The BTZ space-time is given by the
line element \cite{1}
\begin{eqnarray}\label{1}
ds^2=(-f(r)+\frac{16G^2j^2}{r^2})dt^2+\frac{dr^2}{f(r)}+r^2d\varphi^2+8Gjdtd\varphi
\end{eqnarray}
where
\begin{eqnarray}\label{2}
f(r)=(\frac{r^2}{l^2}-8GM+\frac{16G^2j^2}{r^2})
\end{eqnarray}
which is solution of the Einstein equation. The event horizons of
the space-time are given by the singularities of the metric
function which are the real roots of $r^2f(r)=0$. In the above
metric $G$, $j$, $M$ and ~$-\frac{2}{l^2}$ are gravitational
constant, rotational parameter, mass of black hole and
cosmological constant respectively. But their definitions in CTMG
are
\begin{eqnarray}\label{3}
M=m+\frac{1}{\mu}\frac{j}{l^2}~~~~~~~J=j+\frac{1}{\mu}m
\end{eqnarray}
where $\frac{1}{\mu}$ is coupling constant of gravitational
Chern-Simons term. Now we consider a bulk massless scalar field
$\Phi$ propagating in the background of (1). The Klein-
Gordon(KG) equation
\begin{equation}\label{4}
\Box\Phi=\frac{1}{\sqrt{-g}}\partial_{\mu}(\sqrt{-g}g^{\mu\nu}\partial_{\nu})\Phi=0
\end{equation}
can be simplified by assuming the following from of the scalar
field
\begin{eqnarray}\label{5}
\Phi(t,r,\theta,\varphi)=\exp(-i m\varphi+i\omega t) S(\theta)
R(r)
\end{eqnarray}
and for $l=1$  is reduced to the following equation
\begin{eqnarray}\label{6}
[\partial_{u}(\Delta \partial_{u})+\frac{1}{4}[\frac{(\omega
r_{+}-mr_{-})^2}{(u-u_{+})(u_{+}-u_{-})}-\frac{(\omega
r_{-}-mr_{+})^2}{(u-u_{-})(u_{+}-u_{-})}]]R(u)=0
\end{eqnarray}
where
\begin{equation}\label{7}
\Delta=uf(u)~~~~~~~~~u=r^2~~~~~u_{\pm}=r^2_{\pm}
\end{equation}
and
\begin{eqnarray}\label{8}
r_{\pm}=\sqrt{2G(m+j)}\pm\sqrt{2G(m -j)}
\end{eqnarray}
We can show that this equation can be reproduced by the
introduction of conformal coordinates. We introduce the conformal
coordinates \cite{11}.
\begin{equation}\label{9}
\omega^{+}=\sqrt{\frac{u-u_{+}}{u-u_{-}}}\exp(2\pi
T_{R}\varphi+2n_{R}t)~~
\end{equation}
\begin{eqnarray}\label{10}
\omega^{-}=\sqrt{\frac{u-u_{+}}{u-u_{-}}}\exp(2\pi
T_{L}\varphi+2n_{L}t)
\end{eqnarray}
\begin{eqnarray}\label{11}
y=\sqrt{\frac{u_{+}-u_{-}}{u-u_{-}}}\exp(\pi(
T_{R}+T_{L})\varphi+(n_{R}+n_{l})t)~~~
\end{eqnarray}
We define left and right moving vectors by
\begin{eqnarray}\label{12}
H_1=\partial_{+}~~~~
\end{eqnarray}
\begin{eqnarray}\label{13}
H_0=(\omega^{+}\partial_{+}+\frac{1}{2}y\partial_{y}),~~~~ ~
\end{eqnarray}
\begin{eqnarray}\label{14}
~H_{-1}=((\omega^{+})^2\partial_{+}+\omega^{+}y\partial_{y}-y^2\partial_{-})~~~
\end{eqnarray}
\begin{eqnarray}\label{15}
\overline{H}_1=\partial_{-}
\end{eqnarray}
\begin{eqnarray}\label{16}
\overline{H}_0=(\omega^{-}\partial_{-}+\frac{1}{2}y\partial_{y})
\end{eqnarray}
\begin{eqnarray}\label{17}
\overline{H}_{-1}=((\omega^{-})^2\partial_{-}+\omega^{-}y\partial_{y}-y^2\partial_{+})
\end{eqnarray}
which each satisfy the $SL(2,R)$ algebra
\begin{eqnarray}\label{18}
 ~~[H_0,H_{\pm1}]=\mp  H_{\pm 1},~~~~~~~~[H_{-1},H_1]=-2H_0
\end{eqnarray}
and
\begin{eqnarray}\label{19}
~[\overline{H}_0,\overline{H}_{\pm1}]=\mp  \overline{H}_{\pm
1},~~~~~~~~[\overline{H}_{-1},\overline{H}_1]=-2\overline{H}_0
\end{eqnarray}
The quadratic Casimir is
\begin{eqnarray}\label{20}
H^2=\widetilde{H}^2=-H_{0}^2+\frac{1}{2}(H_1H_{-1}+H_{-1}H_{1})=\frac{1}{4}(y^2\partial_{y}^2-y\partial_{y})+y^2\partial_{+}\partial_{-}
\end{eqnarray}
The crucial observation is that these Casimir, when written in
terms  of $ \varphi$ , $t$, $u$
\begin{eqnarray}\label{21}
H^2=\partial_u(\Delta\partial_u)-\frac{u_+-u_-}{u-u_+}(\frac{T_L+T_R}{4A}\partial_t-\frac{n_L+n_R}{4\pi
A }\partial_{\varphi})^2~~~~~~~~\\ \nonumber
+\frac{u_+-u_-}{u-u_-}(\frac{T_L-T_R}{4A}\partial_t-\frac{n_L-n_R}{4\pi
A }\partial_{\varphi})^2~~~~~A=T_L n_R-T_R n_L
\end{eqnarray}
reduces to the radial equation (\ref{6}), with these
identifications
\begin{eqnarray}\label{22}
T_L=\frac{r_++r_-}{2\pi}~~~~~~~~~~~~~T_R=\frac{r_+-r_-}{2\pi}\\
\nonumber n_L=\frac{r_++r_-}{2}~~~~~~~~~~~n_R=\frac{r_--r_+}{2}
\end{eqnarray}
The microscopic entropy of the dual CFT can be computed by the
Cardy formula which matches with the black hole Bekenstein-Hawking
entropy
\begin{eqnarray}\label{23}
S_{CFT}=\frac{\pi^2}3({c_{L}T_{L}+c_{R}T_{R}})~
\end{eqnarray}
The central charges of CTMG are
\begin{eqnarray}\label{24}
c_L=\frac{3}{2G}(1+\frac{1}{\mu})~~~~~~~c_R=\frac{3}{2G}(1-\frac{1}{\mu})
\end{eqnarray}
From the central charges (\ref{24}) and temperature (\ref{22}) we
have
\begin{eqnarray}\label{25}
S_{CFT}=\frac{\pi r_+}{2G}+\frac{1}{\mu}\frac{\pi r_-}{2G}
\end{eqnarray}
which agrees precisely with the gravity result. The contribution
to the entropy that is due to the gravitational Chern-Simons
(last term in eq.(\ref{25})) was first obtained by Solodukhin
\cite{sol}. It is curiously proportional to the area of the inner
horizon rather that of the outer horizon.
\section{Galilean Conformal Algebra in 2-Dimension}
Galilean Conformal Algebra in 2-dimensions can be obtained
 from contracting conformal algebra in 2-dimensions \cite{c}. In two dimensions the non-trivial
generators are given by
\begin{eqnarray}\label{261}
L_n=-(n+1)t^n x\partial_x-t^{n+1}\partial_t, \hspace{1cm}
M_{n}=t^{n+1}\partial_x
\end{eqnarray}
The above generators obey the following commutation relations ,
where define the Galilean conformal algebras.

\begin{eqnarray}\label{271}
 [L_m,L_n]=(m-n)L_{m+n}~~~\\
 \nonumber
 [L_m,M_n]=(m-n)M_{m+n}~\\
 \nonumber
 [M_n,M_m]=0~~~~~~~~~~~~~~~~~~
\end{eqnarray}
2$d$ Conformal algebra at the quantum level are described by two
copy of Virasoro algebra.
\begin{eqnarray}\label{a}
[\mathcal{L}_m,\mathcal{L}_n]=(m-n)\mathcal{L}_{m+n}+\frac{c_{R}}{12}m(m^2-1)\delta_{m+n,0}\\
\nonumber
[\overline{\mathcal{L}}_m,\overline{\mathcal{L}}_n]=(m-n)\overline{\mathcal{L}}_{m+n}+\frac{c_{L}}{12}m(m^2-1)\delta_{m+n,0}
\end{eqnarray}
From these, one obtains centrally extended 2d GCA
\begin{eqnarray}\label{b}
[L_m,L_n]=(m-n)L_{m+n}+C_1m(m^2-1)\delta_{m+n,0}~~\\
\nonumber [L_m,M_n]=(m-n)M_{m+n}+C_2m(m^2-1)\delta_{m+n,0}\\
\nonumber
[M_n,M_m]=0~~~~~~~~~~~~~~~~~~~~~~~~~~~~~~~~~~~~~~~~~~~~~~
\end{eqnarray}
 GCA central charges $C_1$ and $C_2$ are defined in terms
of  CFT central charges \cite{1},
\begin{eqnarray}\label{26}
C_1=\lim_{\epsilon\rightarrow
0}\frac{c_L+c_R}{12}=\frac{1}{4G}~~~~~~C_2=\lim_{\epsilon\rightarrow
0}(\epsilon\frac{c_L-c_R}{12})=\frac{1}{4G\mu}
\end{eqnarray}
CFT entropy (\ref{23}) with the limit ($\epsilon\rightarrow 0$) is
converted to Galilean conformal entropy,
\begin{eqnarray}\label{27}
S_{GCA}=\lim_{\epsilon\rightarrow
0}(\frac{\pi^2}{3}[6C_1(T_L+T_R)+6C_2(\frac{T_L-T_R}{\epsilon})])
\end{eqnarray}
We define Galilean conformal  temperatures as following
\begin{eqnarray}\label{28}
T_1=\lim_{\epsilon\rightarrow 0}
6(T_L+T_R)~~~~~~~~T_2=\lim_{\epsilon\rightarrow
0}6\frac{T_L-T_R}{\epsilon}
\end{eqnarray}
the GCA entropy is
\begin{eqnarray}\label{29}
S_{GCA}=\frac{\pi^2}3({C_1T_{1}+C_{2}T_{2}})~
\end{eqnarray}
To make the GCA entropy finite, $T_L+T_R\sim O(1)$ and
$T_L-T_R\sim O(\epsilon)$, as a result $r_+\sim O(1)$ and
$r_-\sim O(\epsilon)$. These results appear in the limits of
$j\rightarrow \epsilon j$ and $m\rightarrow m$, that correspond
with the result of \cite{1}. Finally from Eq.(\ref{22}) and above
discussion we have introduced finite $n_1$ and $n_2$ (in
non-relativistic limit) in terms of $n_L$ and $n_R$
\begin{eqnarray}\label{30}
n_1=\frac{n_L+n_R}{\epsilon}~~~~~~~~~~n_2=n_L-n_R
\end{eqnarray}
\section{Massless Scalar field in the background of non-relativistic CTMG}
In this section, we will show that for massless scalar field
$\Phi$, in non-relativistic CTMG  there exist a finite  Galilean
conformal symmetry acting on the solution space. We consider a
bulk massless scalar field $\Phi$ propagating in the background
of (1). The KG equation (\ref{6}) in non-relativistic limit
($j\rightarrow \epsilon j$, $\varphi\rightarrow \epsilon
\varphi$) reduce to
\begin{eqnarray}\label{31}
\partial_u(\Delta
\partial_u)-\frac{1}{4}[\frac{r'^{2}_{+}}{(u-u_+)(u-u_-)}\partial_{t}^2~~~~~~~~~~~~~~~~~~~~~~~~~~~~~~~~~~~~~~~\\
\nonumber-2r'_{+}r'_{-}(\frac{1}{(u-u_+)(u-u_-)}-\frac{1}{(u-u_-)(u_+-u_-)})\partial_t\partial_{\varphi}]\Phi=0\\
\nonumber
\partial_{\varphi}^2\Phi=0~~~~~~~~~~~~~~~~~~~~~~~~~~~~~~~~~~~~~~~~~~~~~~~~~~~~~~~~~~~~~~~~~~~~~~~~
\end{eqnarray}
where $r'_{+}=2\sqrt{2Gm}$ and $r'_{-}=\sqrt{\frac{2G}{m}}j$.  We
can show that the above equation can be reproduced by the
conformal coordinate (\ref{9}), (\ref{10}), (\ref{11}) in
non-relativistic limit. $H^2$ is Casimir of $SL(2,R)_{L}\times
SL(2,R)_{R}$ so
\begin{eqnarray}\label{32}
[H^2,H_{0\pm1}]=0~~~~~~~~~~~~[H^2,\overline{H}_{0\pm1}]=0
\end{eqnarray}
Since, $2D$  Galilean conformal generators  are created by the
mixing of  $2D$ conformal generators in non-relativistic limit,
from Eq.(\ref{32}) it can be shown that Casimir of conformal group
is the same  Casimir of Galilean conformal group in
non-relativistic limit. In another term, since
\begin{eqnarray}\label{321}
H'^2=\lim_{\epsilon\rightarrow 0}H^2,
\hspace{0.5cm}L_{0\pm1}=\lim_{\epsilon\rightarrow
0}(H_{0\pm1}+\overline{H}_{0\pm1}),
\hspace{0.5cm}M_{0\pm1}=\lim_{\epsilon\rightarrow
0}(\frac{H_{0\pm1}+\overline{H}_{0\pm1}}{\epsilon})
\end{eqnarray}
we have
\begin{eqnarray}\label{33}
\lim_{\epsilon\rightarrow
0}[H^2,H_{0\pm1}+\overline{H}_{0\pm1}]=0~~~~~~~~~~~~~~\lim_{\epsilon\rightarrow
0}[H^2,\frac{H_{0\pm1}+\overline{H}_{0\pm1}}{\epsilon}]=0\\
\nonumber
[H'^2,L_{0\pm1}]=0~~~~~~~~~~~~~~~~~~~~~~~~~[H'^2,M_{0\pm1}]=0.~~~~~~~~~~~~~~~~~~
\end{eqnarray}
So, the Casimir operator (\ref{21}) in non-relativistic limit is
Casimir of GCA ($H'^2$ is non-relativistic limit of $H^2$).
Finally Casimir operator of GCA is
\begin{eqnarray}\label{34}
H'^2=\partial_u(\Delta\partial_u)-\frac{u_+-u_-}{u-u_-}(\frac{T_1}{A'^2})^2\partial_{t}^2~~~~~~~~~~~~~~~~~~~~~~~~~~~~~~~~\\
\nonumber -\frac{2(u_+-u_-)}{\pi
A'^2}(\frac{T_1n_1}{u-u_+}-\frac{T_2n_2}{u-u_-})\partial_t\partial_{\varphi}~~~~~~~~~~~~~~~~~~~~~~~~~\\
\nonumber
+(\frac{1}{\epsilon^2}\frac{u_+-u_-}{r-r_-}\frac{n_2^2}{\pi^2
A'^2}-\frac{u_+-u_-}{r-r_+}\frac{n_1^2}{\pi^2
A'^2})\partial_{\varphi}^2~~~~~~A'=T_1n_2-T_2n_1
\end{eqnarray}
The above equation is reduced to the radial equation (\ref{31}),
with these identifications.
\begin{eqnarray}\label{35}
T_1=\frac{6r'_+}{\pi}~~~~~~T_2=\frac{6r'_-}{\pi}~~~~~~n_1=r'_-~~~~~~~~n_2=r'_+
\end{eqnarray}
The microscopic entropy of the dual GCA can be computed by the
 non-relativistic Cardy formula (\ref{29}). From the central charges (\ref{26})
and temperatures (\ref{35}) we have
\begin{eqnarray}\label{36}
S_{GCA}=\pi(\sqrt{\frac{2m}{G}}+\frac{j}{\mu}\sqrt{\frac{1}{2Gm}})=S_{BH}
\end{eqnarray}
which agrees precisely with the gravity result, (in
non-relativistic limit) presented in \cite{1}. As we have
mentioned in the introduction the authors of \cite{1} have
studied the Galilean non-relativistic limit of the dual field
theory of BTZ black hole in CTMG. From the Galilean limit of
Virasoro algebra and the Galilean limit of the BTZ black hole,
they get the result that the entropy of the Galilean limit of BTZ
black hole is consistent with the entropy calculated using Cardy
formula from the Galilean limit of the Virasoro algebra. This
agreement between our result and the result of \cite{1} is
interesting, and show that the non-relativistic version of the
BTZ solution of CTMG really has hidden conformal symmetry in the
near region of the black hole. In the next section by obtaining
the absorption cross section we present final evidence of the
existence of this hidden symmetry.

\section{Absorption cross section}
In this section, we give  a brief review for scattering of the
scalar field $\Phi$ which propagates  in the (contracted) BTZ
background \cite{bir}. We calculate the absorption cross section,
from gravity side and matches the result with 2D (GCA) CFT cross
section. Two-point function of conformal invariant fields is
given by \cite{21, mal}
\begin{eqnarray}\label{b1}
G(t^{+},t^{-})\sim (-1)^{h_R+h_L}(\frac{\pi T_L}{\sinh(\pi
T_Lt^{+})})^{2h_L}(\frac{\pi T_R}{\sinh(\pi T_Rt^{-})})^{2h_R}
\end{eqnarray}
where $t^{\pm}$ are the coordinates of the 2D CFT and ($h_R, h_L$)
are eigenvalues of $\mathcal{L}_0$  and $\overline{\mathcal{L}}_0$
respectively. Absorption cross section in term of frequency and
temperature, from Fermi$^{,}$s golden rule \cite{21, mal} can be
read
\begin{eqnarray}\label{b2}
P_{abs}\sim\int dt^{+}dt^{-}e^{-i\omega_R
t^{-}-i\omega_Lt^{+}}[G(t^{+}-i\delta,t^{-}-i\delta)-G(t^{+}+i\delta,t^{-}+i\delta)]
\end{eqnarray}
Using the following integral
\begin{eqnarray}\label{b3}
\int dx e^{(-i\omega x)}(-1)^{\Delta}(\frac{\pi T}{\sinh[\pi
T(x\pm i\delta )]})^{2\Delta}=\frac{(2\pi
T)^{2\Delta-1}}{\Gamma(2\Delta)}e^{\pm\omega/2T}\mid\Gamma(\Delta+i\frac{\omega}{2\pi
T })\mid^2~~~~
\end{eqnarray}
the absorption cross section becomes
\begin{eqnarray}\label{b4}
P_{abs}\sim
T_{L}^{2h_{L}-1}T_{R}^{2h_{R}-1}\sinh(\frac{\omega_{L}}{2T_{L}}+\frac{\omega_{R}}{2T_{R}})\mid\Gamma(h_{L}+i\frac{\omega_{L}}{2\pi
T_{L}})\mid^2\mid\Gamma(h_{R}+i\frac{\omega_{R}}{2\pi
T_{L}})\mid^2
\end{eqnarray}
From above method, non-relativistic limit of absrption cross
section  can be computed. Non-relativistic limit of two-point
function is given by
\begin{eqnarray}\label{b5}
\lim_{\epsilon\rightarrow 0}
G\sim(-1)^{\Delta}(\frac{T_1}{12\sinh(\frac{\pi T_1}{12}(t\pm
i\delta))})^{2\Delta}\exp(\frac{2T_2}{T_1}\xi)
\end{eqnarray}
where scaling dimension $\Delta=h_L+h_R$ is eigenvalue of $L_0$
and rapidity $\xi = \lim_{\epsilon\rightarrow
0}[\epsilon(h_L-h_R)]$ is eigenvalue of $M_0$.  From  Eqs.(
\ref{b2}), (\ref{b3}), (\ref{b5}) and using this relation
\begin{eqnarray}\label{b6}
\lim_{x\rightarrow k}(1+f(x))^{g(x)}=\lim_{x\rightarrow k
}\exp(f(x)g(x))~~~~~~~
\end{eqnarray}
where
\begin{eqnarray}\label{b7}
\lim_{x\rightarrow k}f(x)=0,~~~~~\lim_{x\rightarrow k}g(x)=\infty
\end{eqnarray}
the cross section for 2D GCA is given by

\begin{eqnarray}\label{b8}
P_{abs}\sim\exp(\frac{2T_2}{T_1}\xi)T^{2\Delta-1}_1\sinh(\frac{6\omega_1}{T_1})\mid\Gamma(\Delta+i\frac{6\omega_1}{\pi
T_1 })\mid^2
\end{eqnarray}
where $\omega_1=\omega_L+\omega_R$. We can study, the absorption
cross section of BTZ (contracted BTZ), from gravity side, in
relativistic (non-relativistic) limit. The results can be
verified in agreement with the corresponding results for the
operator dual to the scalar field in the 2D CFT (\ref{b4}) (2D GCA
(\ref{b8})). The radial part of KG equation

\begin{eqnarray}\label{b9}
(\Box-M^2)\Phi=0
\end{eqnarray}
for massive scalar field $\Phi$, is given by
\begin{eqnarray}\label{b10}
[\partial_u(\Delta\partial_u)+(\frac{A}{u-u_+}+\frac{B}{u-u_-}+C)]R(u)=0
\end{eqnarray}
where
\begin{eqnarray}\label{b11}
A=\frac{(\omega r_+ -mr_-)^2}{4(u_+ -u_-)}~~~~~~~~~
B=-\frac{(\omega r_- -mr_+)^2}{4(u_+ -u_-)}~~~~~C=\frac{M^2}{4}
\end{eqnarray}
The wave function satisfying the  ingoing  boundary condition at
the horizon is of the form
\begin{eqnarray}\label{b12}
R(z)=z^{\alpha}(1-Z)^{\beta}F(a,b,c; z)
\end{eqnarray}
where $z=\frac{u-u_+}{u-u_-}$, and

\begin{eqnarray}\label{b13}
\alpha=\sqrt{A}~~~~~~~~~~~~~~~~\beta=\frac{1}{2}(1-\sqrt{1-4C})~~~~~~~~~\gamma=\sqrt{-B}
\end{eqnarray}
where
\begin{eqnarray}\label{b14}
c=1-2i\alpha~~~~~~~~~~~a=\beta+i(\gamma-\alpha)~~~~~~~~~b=\beta-i(\gamma+\alpha)
\end{eqnarray}
The asymptotic form can be read out by taking the limit
$z\rightarrow 1$ and $1-z\rightarrow u^{-1}$
\begin{eqnarray}\label{b15}
R(u)\sim D r^{2h-2}+Er^{-2h}
\end{eqnarray}
where
\begin{eqnarray}\label{b16}
D=\frac{\Gamma(c)\Gamma(2h-1)}{\Gamma(a)\Gamma(b)}~~~~~~~~~~~~~~~E=\frac{\Gamma(c)\Gamma(1-2h)}{\Gamma(c-a)\Gamma(c-b)}
\end{eqnarray} and $h$ is the conformal weight
\begin{eqnarray}\label{b17}
h=\frac{1}{2}(1+\sqrt{1-4C})
\end{eqnarray}
Absorption cross section is captured by the coefficient $D$,
\begin{eqnarray}\label{b18}
P_{abs}\sim \mid
D\mid^{-2}\sim\sinh(2\pi\alpha)\mid\Gamma(a)\mid^2\mid\Gamma(b)\mid^2
\end{eqnarray}
To see explicitly that $P_{abc}$ matches with microscopic
greybody factor of the dual CFT we need to identify the conjugate
charges, $\delta E_L$ and $\delta E_R$ defined by
\begin{eqnarray}\label{b19}
\delta S_{BH}=\delta S_{CFT}=\frac{\delta E_L}{T_L}+\frac{\delta
E_R }{T_R}
\end{eqnarray}
we have
\begin{eqnarray}\label{b20}
\delta E_L=\omega_L~~~~~~~~~~\delta E_R=\omega_R
\end{eqnarray}
where
\begin{eqnarray}\label{b21}
\omega_R=\frac{\omega+m}{2(r_++r_-)}~~~~~~~~~\omega_L=\frac{\omega-m}{2(r_+-r_-)}
\end{eqnarray}
Hence, the coefficient $a, b$ can be expressed in terms of
parameters $\omega_L$ and $\omega_R$
\begin{eqnarray}\label{b22}
a=h_R+i\frac{\omega_R}{2\pi
T_R}~~~~~~~~~~~b=h_L+i\frac{\omega_L}{2\pi T_L}
\end{eqnarray}
where for the scalar field $h_L=h_R=h$. Similarly we have
\begin{eqnarray}\label{b23}
2\pi\alpha=\frac{\omega_L}{2T_L}+\frac{\omega_R}{2T_R}
\end{eqnarray}
Finally from Eqs.(\ref{b18}, \ref{b22}, \ref{b23}), the
absorption cross section can be expressed as
\begin{eqnarray}\label{b24}
\nonumber P_{abs}\sim
T_{L}^{2h_{L}-1}T_{R}^{2h_{R}-1}\sinh(\frac{\omega_{L}}{2T_{L}}+\frac{\omega_{R}}{2T_{R}})\mid\Gamma(h_{L}+i\frac{\omega_{L}}{2\pi
T_{L}})\mid^2\mid\Gamma(h_{R}+i\frac{\omega_{R}}{2\pi
T_{L}})\mid^2
\end{eqnarray}
which is the finite temperature absorption cross section for 2D
CFT. In  following, we consider the scattering of
non-relativistic limit of BTZ, from gravity side, and matches the
result with 2D GCA cross section (\ref{b8}). The radial part of KG
equation in non-relativistic limit can be expressed in term of
Eq.(\ref{b10}), with these  identifications
\begin{eqnarray}\label{b25}
\nonumber A_n=\frac{(\omega r'_+ -mr'_-)^2}{4u'_+ }~~~~~~~~~
B_n=-\frac{(\omega \epsilon r'_-
-mr'_+/\epsilon)^2}{4u'_+}~~~~~C=\frac{M^2}{4}
\end{eqnarray}
From Eqs.(\ref{b13}, \ref{b14}), we  see that
\begin{eqnarray}\label{b25}
\lim_{\epsilon\rightarrow
0}a=a_n=\infty,~~~~~~~~~~~~~\lim_{\epsilon\rightarrow
0}b=b_n=\infty ,~~~~~~~~\lim_{\epsilon\rightarrow 0}(a+b)=finite
\end{eqnarray}
so coefficient $D$ in non-relativistic limit is given by
\begin{eqnarray}\label{b27}
D_n=\frac{\Gamma(c_n)\Gamma(2h-1)}{\Gamma(a_n+b_n)}
~~~~~~~~~~\lim_{a\rightarrow\infty}\Gamma(a)\lim_{b\rightarrow\infty}\Gamma(b)\simeq\Gamma(a_n+b_n)
\end{eqnarray}
Absorption cross section from gravity side in the non-relativistic
limit, is captured by coefficient $D_n$
\begin{eqnarray}\label{b28}
P_{abs}\sim \mid
D_n\mid^{-2}\sim\sinh(2\pi\alpha_n)\mid\Gamma(a_n+b_n)\mid^2
\end{eqnarray}
To see explicitly that $P_{abc}$ matches with microscopic
greybody factor of the dual GCA we need to identify the conjugate
charge, $\delta E_1$  defined by
\begin{eqnarray}\label{b29}
\lim_{\epsilon\rightarrow 0}\delta S_{BH}=\delta
S_{GCA}=\frac{\delta E_1}{T_1}
\end{eqnarray}
we have
\begin{eqnarray}\label{b30}
\delta E_1=\omega_1=\lim_{\epsilon\rightarrow 0
}(\omega_L+\omega_R)
\end{eqnarray}
where
\begin{eqnarray}\label{b31}
\omega_1=\frac{\omega r'_+-mr'_-}{2r'_+}
\end{eqnarray}
Hence, the coefficients $a_n, b_n$ can be expressed in term of
parameters $\omega_L$ and $\omega_R$
\begin{eqnarray}\label{b32}
a_n+b_n=\Delta+i\frac{6\omega_1}{\pi T_1}~~~~~~~~~~~
\end{eqnarray}
where for the scalar field $\Delta=2h$. Similarly we have
\begin{eqnarray}\label{b33}
2\pi\alpha_n=\frac{6\omega_1}{T_1}
\end{eqnarray}
Finally from Eqs.( \ref{b28}), (\ref{b32}), (\ref{b33}), the
absorption cross section can be expressed as ($\xi=\epsilon
(h_L-h_R)=0$)
\begin{eqnarray}\label{b34}
 P_{abs}\sim
T^{2\Delta-1}_1\sinh(\frac{6\omega_1}{T_1})\mid\Gamma(\Delta+i\frac{6\omega_1}{\pi
T_1 })\mid^2
\end{eqnarray}
which is the  absorption cross section for 2D GCA.
\section{Conclusions}
In this paper at first we showed that there exist a holographic
$2D$ CFT description for BTZ black hole solution of CTMG. The key
ingredient for this is the hidden conformal symmetry in the black
hole. We considered the wave equation of a massless scalar field
propagating in CTMG spacetime and found the wave equation could
be written in terms of the $SL(2,R)$ quadratic Casimir. We read
the temperatures of the dual CFT from conformal coordinates. We
recovered the macroscopic entropy from the microscopic counting.
Then we extended this study to the non-relativistic limit, and
showed that there exist a correspondence between $GCA_2$ on the
boundary and contracted BTZ in the bulk. By definition Galilean
conformal temperatures as equation (\ref{28}), we could obtain
the finite $GCA_2$ entropy by equation (\ref{29}). After that we
showed that the radial part of Klein-Gordon (KG) equation in the
background of contracted BTZ black hole, where ($j\rightarrow
\epsilon j$, $\varphi\rightarrow \epsilon \varphi$), can be given
by the non-realtivistic limit of Casimir $H^2$. We could read the
$GCA_2$ temperatures $T_1$, $T_2$ from the correspondence of
radial part of non-relativistic KG equation and GCA Casimir
$H'^{2}$. The temperatures $T_1$, $T_2$ given by equation
(\ref{35}) are exactly equal with previous formula we have obtain
in (\ref{28}). Then we calculated the entropy of contracted BTZ
black hole by using GCA Cardy formula and non-relativistic
temperatures. Finally we have shown that the absorption cross
section of a near-region scalar field match precisely to that of
microscopic dual $GCA$ side.

\section{Acknowledgments} We acknowledge helpful discussions with Takahiro
Nishinaka, and Arjun Bagchi. Also we thank Sergey Solodukhin for
suggesting former reference for Holography with Gravitational
Chern-Simons term.

\end{document}